# Collection of Historical Weather Data: Issues with Missing Values

Fadoua Rafii
UCD, University College Dublin
Belfield, Dublin 4, Ireland
fadoua.rafii@ucd.ie

Tahar Kechadi
UCD, University College Dublin
Belfield, Dublin 4, Ireland
tahar.kechadi@ucd.ie

## ABSTRACT
Weather data collected from automated weather stations have become a crucial component for making decisions in agriculture and in forestry. Over time, weather stations may become out-of-order or stopped for maintenance, and therefore, during those periods, the data values will be missing. Unfortunately, this will cause huge problems when analysing the data. The main aim of this study is to create high-quality historical weather datasets by dealing efficiently with missing values. In this paper, we present a set of missing data imputation methods and study their effectiveness. These methods were used based on different types of missing values. The experimental results show that two of the proposed methods are very promising and can be used at larger scale.

## Keywords
Temperature; Rainfall; Missing data; Imputation methods; RMSE.

## 1. INTRODUCTION
Weather data is gaining popularity within precision agriculture community, however, its collection, storage, and analysis present huge concerns in relation to its quality and usage [1]. Weather is one of the key sources that affects directly crop yield, particularly in the context of climate change [2] [3] [4]. The accurate modelling and the use of modern data analytics approaches will help the farmers to make efficient decisions about what and when they farm, how to manage the fields, and when to drill and harvest [5]. One of the challenges of historical weather data collected over a long period of time is the missing data values. In a time series environment, handling missing data represents a serious issue in forecasting [6]. Missing data are not only causing difficulties in estimating parameters and identifying processes but can be the cause of misinterpretations regarding the temporal and the spatial variations of environmental indicators [7]. Incomplete data in a historical dataset is considered as an unavoidable problem when dealing with the real-world data sources [8]. This issue has been analysed and discussed thoroughly in the literature [9]. Among the relevant weather variables, temperature and rainfall are the key factors in the crop yield analysis [10] [11]. Temperature influences length of growing season and rainfall affects the production of plant [12] [13]. These two weather variables must be included in any agricultural production prediction model.

Rainfall analysis plays an important and significant role in the field of agriculture and ecological studies [14] [15] [16] [17]. The presence of missing values in rainfall data is a common problem in the process of data analysis [18]. In the literature, many authors suggested various imputation methods for estimating rainfall missing values. Little and Rubin [19] concluded that the performance of any imputation method depends on several weather factors:
- Nature of occurrences
- Neighbouring stations
- Intrinsic characteristics

For estimating missing rainfall values, Suhalia et al. [20] and Silva et al. [21] tested multiple methods such as normal ratio, arithmetic mean, inverse distance, aerial precipitation ratio, correlation ratio method, inverse weighting distance and correlation coefficient methods. To compare these methods, they used techniques such as Mean Absolute Error (MAE), correlation coefficient (R) and Similarity index (S index). The Normal Ratio method (NR) is the most common method used for estimating missing rainfall data [22].

Temperature could be used for controlling many physical and biological processes between atmosphere and Earth surface, including transpiration, respiration, and photosynthesis [23]. For imputing temperature missing values, the most accurate method depends on data interpolation schemes [24] [25]. R. P. De Silva et al. [26] compared the Normal Ratio method to Arithmetic Mean, Aerial Precipitation Ratio (APR), and Inverse Distance methods. They concluded that the normal ratio method is the most suitable method compared to the three others.

Yozgatligil et al. [27] compared several imputation methods for completing the missing values in spatiotemporal meteorological time series. They artificially created missing data in monthly mean temperature and total precipitation obtained from the Turkish State Meteorological Service. One of the techniques used is the normal ratio which produced more robust and better results than simple arithmetic average and normal ratio weighted with correlations.

## 2. METHODLOGY
### 2.1 Dataset
The data is collected from twelve temperature and rainfall measuring stations that were selected based on the coverage areas. These areas are located in the UK. The weather data was recorded over a 5-year period (2014 to 2018). The data values were recorded every fifteen minutes.

**Table 1. Records number for each year**

| Year | Days number | Records number |
|---|---|---|
| 2014 | 365 | 35040 |
| 2015 | 365 | 35040 |

| 2016 | 366 | 35136 |
| 2017 | 365 | 35040 |
| 2018 | 295 | 28320 |

Table 1 lists the records number that we should have for each year. As we started analysing the data by the fourth quarter of 2018, we have 295 days of the same year. Table 2 contains geographical coordinates of the nearest weather stations of each weather station. Table 3 presents the maximum and minimum temperatures for the four weather stations.

**Table 2. Geographical coordinates of nearest weather stations**

| Target station | Nearest station | Longitude | Latitude |
|---|---|---|---|
| Station 1, Wales | 1st nearest station | -3.315 | 51.128 |
| | 2nd nearest station | -2.924 | 51.308 |
| Station 2, Central | 1st nearest station | -1.524 | 52.057 |
| | 2nd nearest station | -1.089 | 51.715 |
| Station 3, South-West | 1st nearest station | -2.882 | 50.934 |
| | 2nd nearest station | -2.924 | 51.308 |
| Station 4, West-Midlands | 1st nearest station | -1.524 | 52.057 |
| | 2nd nearest station | -2.003 | 51.703 |

**Table 3. Maximum and minimum temperatures for the target weather stations**

| Target station | Temperature maximum | Temperature minimum |
|---|---|---|
| Station 1, Wales | 59.1 | -6.0 |
| Station 2, Central | 55.7 | -29.9 |
| Station 3, South-West | 60.0 | -35.2 |
| Station 4, West-Midlands | 31.4 | -32.8 |

## 2.2 Analysis Techniques

The purpose of this study is to select an appropriate technique for estimating missing values for the variables; temperature and rainfall. We chose a technique based on revised normal ratio methods [28] and showed a good performance improvement compared to normal ration methods. This approach has been used for the temperature and extended to rainfall. For the evaluation purposes, we implement various approaches with the view to compare to our results.

### 2.2.1 Normal Ratio Method

The normal ratio (NR) method was firstly suggested by Paulhus and Kohler in 1952 [29] and then it was updated by Young in 1992 [22]. It is based on mean ratio of data between a target station and neighbouring stations [28]. There is another version of normal ratio method that is called old normal ratio (ONR). It was used for estimating missing rainfall records [30]. The weighting factor for ONR is arithmetic mean. The main equation of this method is as follows:

$$Y_s = \frac{1}{T}\sum_{i=1}^{T}\left(\frac{M_s}{M_i}\right)Y_i \quad (1)$$

where $Y_s$ is the missing value of temperature or rainfall at a target station; $T$ is the number of nearest (neighbouring) stations; $M_s$ is the sample mean of available data at a target station; $M_i$ is the sample mean of available data at the i$^{th}$ neighbouring station and $Y_i$ is the observed value of temperature or rainfall at the i$^{th}$ neighbouring station.

### 2.2.2 Geographical Coordinates Method

Geographical coordinate (GC) method is a weighting method which is used for imputing missing rainfall values [28]. It uses the inverse of geographical coordinates (latitude and longitude) to calculate weight coefficient. In GC method, the centre point represents a target station. The distance from a centre point to surrounding stations is computed in order to determine the nearest stations.

$$Y_s = \sum_{i=1}^{N}\left(\frac{\frac{1}{x_i^2+y_i^2}}{\sum_{i=1}^{N}\left(\frac{1}{x_i^2+y_i^2}\right)}\right)Y_i \quad (2)$$

where $Y_s$ is the missing value of temperature or rainfall at a target station; $x_i$ is longitude of the i$^{th}$ neighbouring station; $y_i$ is latitude of the i$^{th}$ neighbouring station and $Y_i$ is the observed value of temperature or rainfall at the i$^{th}$ neighbouring station.

### 2.2.3 Normal ratio with geographical coordinates method

This method (NRGC) consists on combining both methods NR and GC mentioned in this study. Since NR is often found as a best estimation method, it adapts the location element in order to upgrade performance. The equation is expressed as follows:

$$Y_s = \sum_{i=1}^{N}\left(\frac{\left(\frac{1}{x_i^2+y_i^2}\right)\left(\frac{M_s}{M_i}\right)}{\sum_{i=1}^{N}\left(\frac{1}{x_i^2+y_i^2}\right)\left(\frac{M_s}{M_i}\right)}\right)Y_i \quad (3)$$

### 2.2.4 Nearest Neighbour Method

One of the simple methods for filling missing values is nearest neighbour (NN). It consists on taking a nearest neighbouring station and using its observation to fill in some missing values in the local station [31]. The selection of a nearest neighbour can be done geometrically or by taking the station that has highest correlation with the target location. The value of a nearest neighbouring station can be transferred directly without making any change [32]. In the literature, we found other methods based on similar concept such as Hot deck imputation which consists on identifying the most similar case to the case that has a missing value and substituting the most similar case's X value for the missing case's X value [8]. There is also Closest station method that Wallis et al. [32] used, where they combined a variant of the hot-deck infilling method with the mean value infilling method. In this study, they identified the closest three stations, and the missing days were estimated from closest station with data. If a closest neighbouring station does not have data, then data are token from a next nearest neighbouring station. If none of the three closest stations have data, then the infilling value is a long-term mean for an appropriate base station and month. In addition, there is Single best estimator (SBE) method which is an analogous method to the fact of using closest neighbouring station to fill gaps for a target station [33].

## 3. EXPERIMENTAL RESULTS
### 3.1 System Settings

The process for analysing weather data is shown in Figure 1.

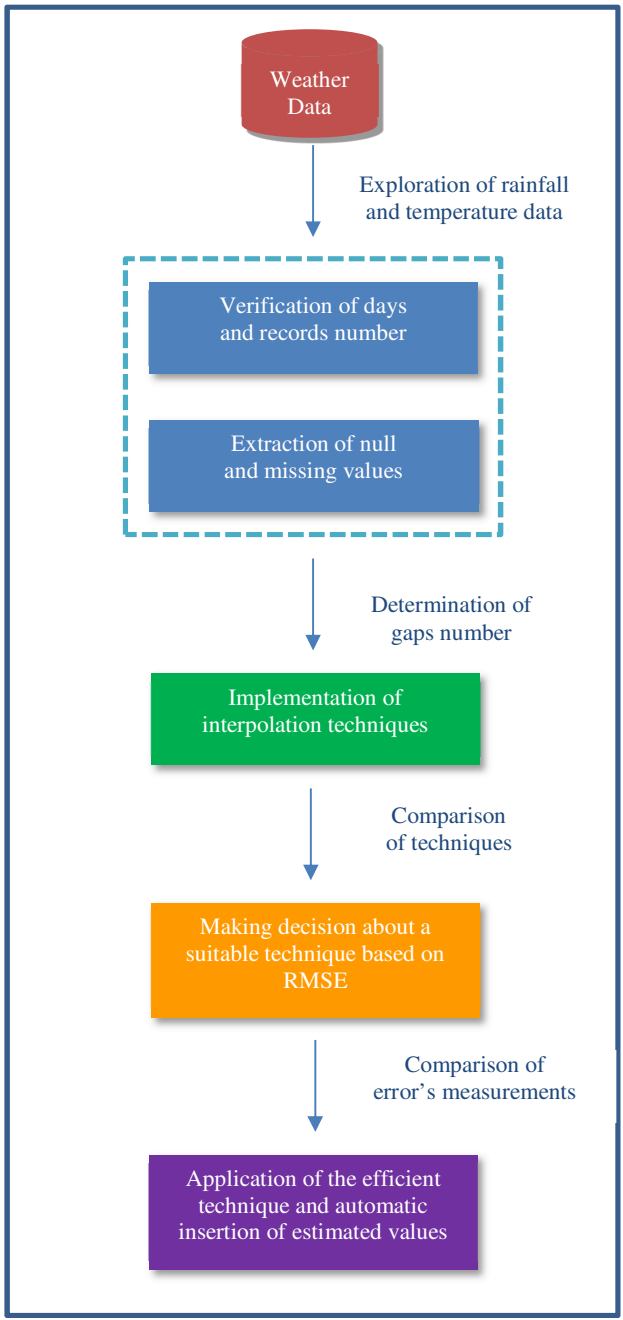

**Figure 1. Workflow for analyzing weather stations data.**

The first step consists on exploring and focusing on rainfall and temperature data. In the second step, we developed an algorithm to validate the number of days, number of records, null and missing values for each specific year and for each weather station. The third step looks for the number of gaps (interval) of missing values or null values that are less than one hour. In this case, we use a simple interpolation by taking into consideration previous and next values. In the case where the gaps are bigger than one hour, we use an interpolation technique that takes into account the neighbouring stations. In the fifth step, based on different results generated by interpolation techniques and depending on RMSE, a decision is made for others weather stations. The final step is to fill gaps and process null values automatically with estimated rainfall and temperature values.

## 4. RESULTS

For comparing the four implemented techniques, deviation or error between an actual and an estimated value should be calculated. The Root Mean Square Error is considered a representative value for measuring errors. The four imputation methods were compared to each other's.

Concerning rainfall data, Figure 2 shows the results obtained by implementing the techniques on four weather stations. It depicts clearly the efficiency of methods based on geographical coordinates. On the other hand, on Figure 3 both methods NRGC and GC produced small RMSE error. Besides, NN method produced good accuracy for the four stations.

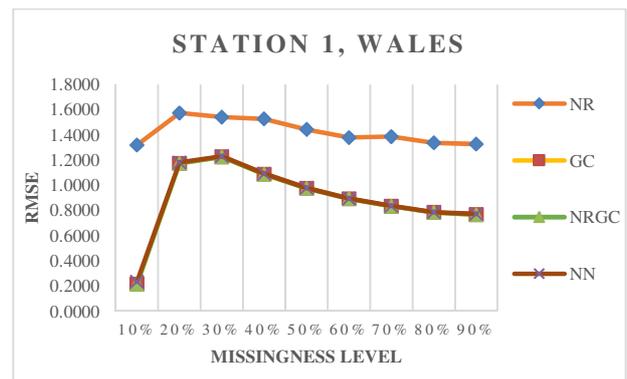

(a)

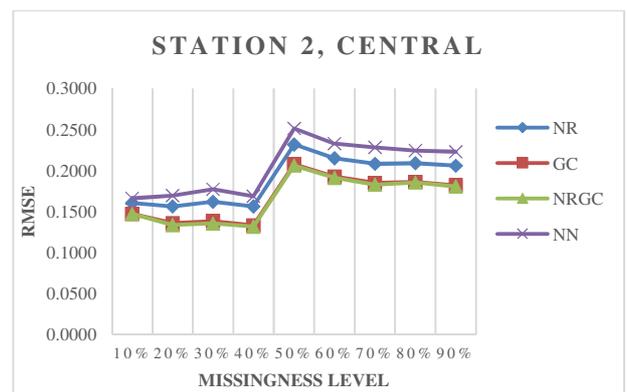

(b)

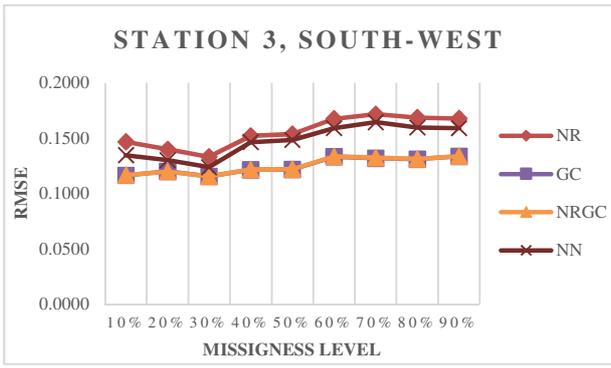

(c)

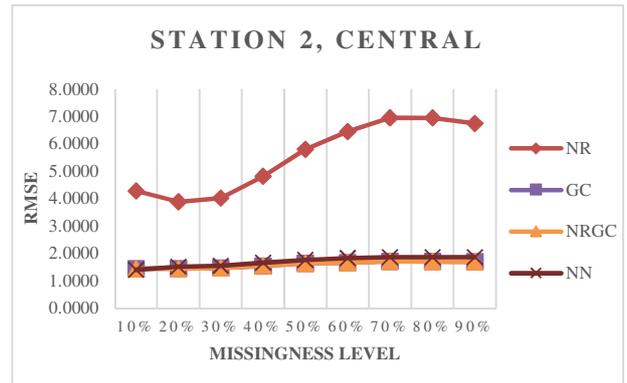

(f)

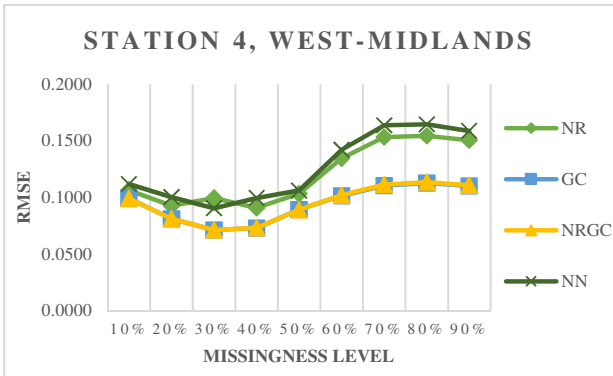

(d)

**Figure 2. Performances of imputation methods at different missingness levels of rainfall data based on RMSE for Station 1 Wales (a), Station 2 Central (b), Station 3 South-West (c) and Station 4 West-Midlands (d)**

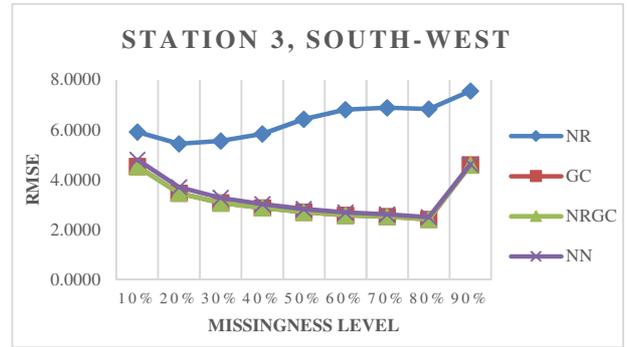

(g)

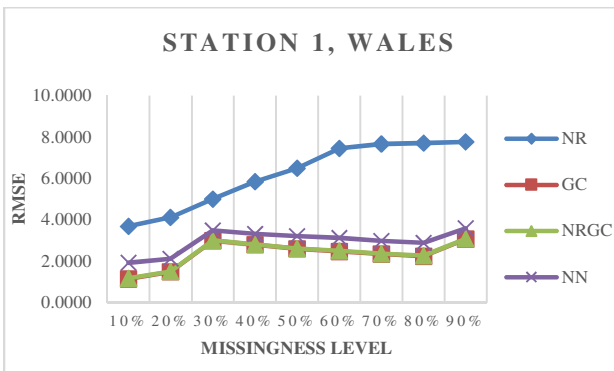

(e)

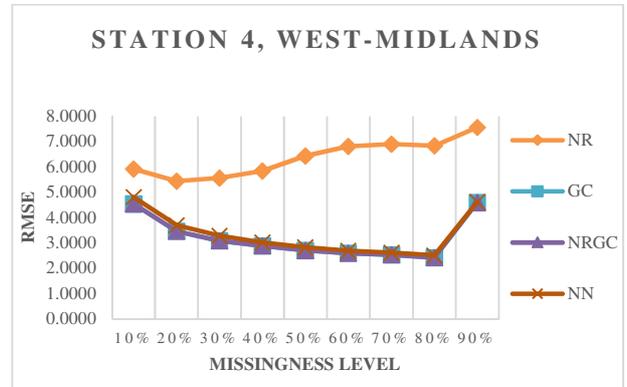

(h)

**Figure 3. Performances of imputation methods at different missingness levels of temperature data based on RMSE for Station 1 Wales (e), Station 2 Central (f), Station 3 South-West (g) and Station 4 West-Midlands (h)**

## 5. CONCLUSION

Good quality temperature and rainfall data are necessary for agricultural decision support. In this study, we aimed to explore the estimation methods in order to choose the most accurate one for both temperature and rainfall missing values. The performance of the four methods are tested and evaluated using state of the art measures. Based on these measures, we have shown that NRGC and GC methods are the most accurate and appropriate for estimating the temperature and rainfall missing values. Due to the unavailability of data (privacy issues), we were concentrated on UK. We will generalize the results by implementing the same techniques on other real data from different countries.

## 6. ACKNOWLEDGMENTS

This research forms part of the CONSUS Programme which is funded under the SFI Strategic Partnerships Programme (16/SPP/3296) and is co-funded by Origin Enterprises Plc.